\documentclass[aps,twocolumn,prd,showpacs]{revtex4-1}
\usepackage{graphicx}

\usepackage{amsfonts}
\usepackage{amssymb}

\usepackage{array}
\usepackage{graphicx}
\usepackage{pstricks}
\usepackage{color}
\usepackage{xcolor}

\usepackage{subfigure}
\usepackage{amsmath}

\usepackage{latexsym}

\begin{document}
\title{Geometric phase of neutrino propagating through dissipative 
matter}\email[~Published~in: Phys.Rev.D {\bf 83}, 097302 (2011); \\
DOI: 10.1103/PhysRevD.83.097302]{}

\author{ J. Dajka}
\author{J. Syska}
\author{J. {\L}uczka  }
\affiliation{Institute of Physics, University of Silesia, 40-007
Katowice, Poland }

\begin{abstract}
We study the geometric phase (GP) in neutrino oscillation for both Dirac and Majorana neutrinos. We
apply the kinematic generalization of the GP to quantum open systems that take into account the coupling
to a dissipative environment. In the dissipationless case, the GP does not depend on the Majorana angle. It is not the case in the presence of dissipation and hence the GP can serve as a tool determining the type of the Dirac vs the Majorana neutrino.
\end{abstract}


\maketitle
%

The physics of neutrino has inspired the long-standing  debate, at least in two clearly recognizable issues. The first
is related to the existence of the  neutrino mass
\cite{Lisi-Marrone-Montanino,bf1}. The 
second concerns the nature of the Dirac vs the Majorana
neutrino. 
Recent studies suggest that the subtle, 
quantum   phenomenon of the neutrino interference \cite{neu_interf,Mehta} gives further
insight into this second issue \cite{Bilenky_Doi,AZS_ZZS}.
This is due to the geometric phase (GP),
 the property of quantum evolution
already recognized as a hallmark of various neutrino features \cite{neu_faza}.
The concept of the GP has been elucidated in various context of classical and quantum physics, including quantum information with a potential application in holonomic quantum computation as a means of constructing built-in fault tolerant quantum
logic gates.
We propose to  study the GP  for neutrinos
to achieve two aims at once: (1) to compare its behavior for an ideal closed system and  an experimentally more realistic open system in  the presence of matter, noise, and environments, and (2)  to  exploit its properties to distinguish between the Dirac and Majorana neutrinos.
Although one should consider  three neutrino flavors for a complete analysis, within this work we
study a minimal model based on two neutrino flavors.  In case of electron-muon neutrino oscillation  this approximation is  fairly justified due to hierarchy of mass splittings and small values of a part of elements of the  mixing matrix, the fact we refer below.
It is effectively described  as a {\it two-level} system with a suitably defined Hamiltonian. Recently, it   has been  applied in the context of the entanglement dynamics  \cite{neu_ent}.
In this paper, using
the model of the dissipative Markovian dynamics which includes  effects of both deterministic and noisy interactions between the neutrino and the ordinary matter \cite{bf1}, we analyze how the GP acquired by the oscillating neutrino in the (quasi)--cyclic evolution indicates whether it is the Dirac  fermion or Majorana one. 

Let us consider  two neutrino flavors,
the electron $(e)$ and muon ($\mu$) one with  two orthogonal vacuum massive
states $|1\rangle $ and $|2\rangle $.
This approach is useful for solar experiments under the
experimental settings of {\it the active} $\Delta m_{21}^{2}$ \cite{Giunti-Kim}:  $(1/2) \Delta m_{21}^{2} c^3 \langle L/ \hbar E\rangle \sim \pi$ and $2 E V_{C} \sim \Delta m_{21}^{2} c^{4}$, where $E$ and $L$ are the energy of the massless neutrino and experimental baseline, respectively. Here $\Delta m_{21}^{2} \equiv m_{2}^{2} - m_{1}^{2}$ is the square mass splitting of the mass states $i=1,2$
in the normal hierarchy case,  $ V_{C} = \sqrt{2} \, G_{F} N_{e}$
is the effective  potential of the neutrino in the ordinary matter
due to the coherent forward scattering on electrons via charged current (CC) interactions, where $N_{e}$ is the electron density
and $G_{F}$ is the Fermi constant \cite{Giunti-Kim}.
Massive states $|1\rangle $ and $|2\rangle $
are associated with flavor ones: the electron neutrino state
$|\nu_e \rangle = \cos \theta \,|1\rangle  + e^{i \phi} \sin \theta \,|2\rangle $ and the muon one
$|\nu_\mu\rangle = -  \sin \theta \,|1\rangle  +  e^{i \phi} \cos \theta \,|2\rangle $, where $\theta = \theta_{12}$ is the mixing angle and $\phi$ is the ($CP$-violating) Majorana phase.
In the Dirac neutrino case,
$\phi$ can be eliminated via the $U(1)$ global transformation,  $|1\rangle  \rightarrow  |1\rangle $ and $|2\rangle  \rightarrow e^{-i \phi}  |2\rangle $ \cite{Giunti-Kim}. In turn, in the Majorana 
case, the mass term in the Lagrangian is not invariant under the above transformation and
rephasing of the  left-chiral massive neutrino  field
is not possible, leaving
$\phi$ nonzero.  However, it does not contribute to  oscillation formulas in the standard model with nonzero neutrino mass ($\nu$SM) \cite{Giunti-Kim}.

The corresponding  initial density matrix  for the electron neutrino  reads  \cite{bf1}:
\begin{eqnarray}\label{init}
\rho ^{e}(0)&=&\left(\begin{array}{cc} \cos^2 \theta & \frac{1}{2}\mbox{e}^{-i\phi}\sin 2\theta  \\ \frac{1}{2}\mbox{e}^{i\phi}\sin 2\theta  & \sin^2 \theta \end{array}  \right) \; ,
\end{eqnarray}
and for the muon neutrino
$\rho ^{\mu}(0) = 1 - \rho ^{e}(0)$.

From now on  we assume  that the neutrino propagates through  matter and interacts with its environment.
It is a source of decoherence and dissipation which allows transitions from  the pure  state to mixed one. In the presence of dissipation,
the Majorana phase can enter both into the transition probabilities \cite{bf1,neu_faza,neu_ent} and  the neutrino geometric phase.
For the GP to be detected it is necessary to perform the split-beam-interference  experiment.
%
%
As the neutrino cross section is very tiny,
%
%
until now the spatially beam splitting experiment is impossible.
However, the flavor neutrino is the superposition of  two massive states
 which splits just at the moment of the production of its $\alpha$-flavor superposition; then  it propagates and finally   two massive states interfere in the detector in its $\beta$-flavor interference pattern. This   single flavor neutrino
split-beam experiment in the energy space is the one we need  \cite{Mehta}. \\

In what follows we suppose that
the neutrino is the relativistic particle
hence
the vacuum mass states $|i \rangle$
have energies $E_{i} \approx E + m_{i}^{2}c^{4}/2 E$, $i=1,2$~ \cite{Giunti-Kim}.
Then the neutrino vacuum Hamiltonian is
\begin{eqnarray}
\label{H effective vaccum}
H_{0}  = \left(
\begin{array}{cc}
 E - \frac{\Delta m_{21}^{2}c^4}{4 E}& 0 \\
 0 & E + \frac{\Delta m_{21}^{2}c^4}{4 E}
\end{array}
\right) \;   \;
\end{eqnarray}
and the neutrino Wolfenstein effective Hamiltonian $H$
in medium,  in the vacuum neutrino mass basis \cite{bf1,Giunti-Kim} is
\begin{eqnarray}
\label{Heff}
H  &=& H_{0}  + \frac{V_0}{2} \left(
\begin{array}{cc}
   1 + \cos 2 \theta &  e^{- i \phi} \sin 2 \theta \\
  e^{i \phi} \sin 2 \theta &
 1 -  \cos 2 \theta
\end{array}
\right),
\end{eqnarray}
 where  $V_{0} = \cos^{2}(\theta_{13}) V_{C}$ is the interaction potential
and $0.953 < \cos^{2}\theta_{13} \leq 1$ is one of the oscillation parameters with $3 \sigma$ bound  \cite{Giunti-Kim}.
We take into account the usual matter only hence the corresponding CC 
term for the muon neutrino is missing.
In the $\nu$SM  the neutral current interaction
does not enter effectively into  Eq.(\ref{Heff})  \cite{Giunti-Kim}. 

Because neutrino propagates in matter and interacts with
its environment leading to decoherence and dissipation
hence the considered system should be treated as an open
system, which in the Markovian regime can be described
by completely positive linear maps acting on the system
density matrices. Their general form reads  \cite{Alicki}
\begin{eqnarray}
\label{mast}
\frac{d}{dt} \rho^{\alpha} (t) = - i [H, \rho^{\alpha} (t)] + L[\rho^{\alpha} (t)], \;\;
\alpha = \{e, \nu\}.
\end{eqnarray}
One can recognize in (\ref{mast}) the Kossakowski--Lindblad master equation with two parts responsible for the physically distinct processes. The first
(conservative) part is generated by the effective Hamiltonian  $H$. The second (dissipative) part is generated by the dissipator $L$ and results in the nonunitary evolution of the density matrix. If one knows all details of the system-environment iteration, it is possible (in principle) to construct the corresponding dissipator.
The (semi)phenomenological treatment of the  neutrino propagation with the dissipation is presented in \cite{bf0,bf1} with the dissipator in the form 
\begin{eqnarray}
\label{dys}
L[\rho^{\alpha} ] &=& \sum_{i,j=0}^3 C_{ij} \left(\sigma_j \, \rho^{\alpha}  \, \sigma_{i} - \frac{1}{2} \left\{\sigma_{i} \sigma_{j} ,\, \rho^{\alpha}  \right\} \right),
\end{eqnarray}
where $\sigma_i$ are the Pauli matrices and $C_{ij}$ should assure  complete positivity of the map.  
The constraints guaranteeing complete positivity applied  to $C_{ij}$ result in reducing the number of free parameters to six as discussed in \cite{bf0,bf1}. Here, instead of attempting to derive the relation between $C_{ij}$ and the properties of an  environment \cite{bf1}, we  consider the dissipator (\ref{dys}) as a result of a phenomenological modeling. Such an approach is clearly less physical, as it suffers from a lack of microscopic justification. On the other hand, phenomenological modeling guided exclusively by the requirement of complete positivity remains independent on any approximation always used in more fundamental derivations.   Let us notice that the effective description of nonstandard effects resulting from openness of the system has recently been applied to various systems in particle physics \cite{bf_inne}.

There have been many proposals tackling the problem of the geometric phase from different generalizations of the parallel transport condition for systems which are either in a mixed
state and/or undergo a nonunitary evolution like that determined by Eq. (\ref{mast}). The earliest attempt (purely mathematical) towards
this goal is given in \cite{armin}.  The others are   based on
quantum trajectories \cite{traj}, quantum interferometry \cite{sjuk1}, and the state
purification (kinematic approach) \cite{sjuk2}.
Here we use the kinematic approach. The GP constructed in \cite{sjuk2}
exhibits primary features: it  is  purification-independent, gauge invariant and reduces to the standard definition  in the limit of an unitary evolution.
 One of the appealing advantages of studying
this  phase  is its measurability in   a
carefully prepared interferometric experiments \cite{sjuk1,sjuk2}. 
A new type of an experiment on the GP of open systems has recently been reported \cite{fazmeasure}: 
the GP has been determined by measuring the decoherence factor
of the off-diagonal elements of the reduced density matrix
of the system. 
Our reasoning is thus guided by its potential for experimental
implementation.
 In order to determine  the GP based on
state purification \cite{sjuk2} we have to  rewrite the
density matrix   in the spectral-decomposition   form
\begin{eqnarray}
\label{spect}
\rho^{\alpha}(t)=\sum_{i=1}^{2} \, \lambda_{i}^{\alpha}(t) \, |w_{i}^{\alpha}(t)\rangle \langle w_i^{\alpha}(t)|,
\end{eqnarray}
where
$\lambda_{i}^{\alpha}(t)$ and $|w^{\alpha}_{i}(t)\rangle$ are the   instantaneous eigenvalues and the eigenvectors of the matrix $\rho^{\alpha}(t)$,
respectively. The GP $\Phi^{\alpha}(t)$ corresponding to such an
evolution  is defined by the relation \cite{sjuk2}:
\begin{eqnarray} \label{Phi}
\Phi^{\alpha}(t) &=& \mbox{Arg}\left[\sum_{i=1}^2 [\lambda_i^{\alpha}(0) \lambda_i^{\alpha}(t)]^{1/2}\langle
w^{\alpha}_i(0) |w^{\alpha}_i(t)\rangle\right.\nonumber\\ && \left. \times
\exp(-\int_0^t \langle w^{\alpha}_i(s)|\dot{w}^{\alpha}_i(s)\rangle ds)\right],
\end{eqnarray}
where  $\mbox{Arg}\left[z\right]$ denotes argument  (or phase) of the complex number $z$,  $\langle w^{\alpha}_i|w^{\alpha}_j \rangle$ is a scalar product and   the dot indicates the
derivative with respect to time $s$. Below we  consider
 the electron neutrino  only \cite{Giunti-Kim} so, $\Phi(t)= \Phi^e(t)$.  

For the closed,  dissipationless system in vacuum ($V_0=0$), the evolution of the neutrino is unitary and cyclic with the period $T= L/c=\pi/\omega_0$, where
$\omega_0 = \Delta m_{21}^{2} c^4/4\hbar E$. In this case, the GP assumes the well-known form
\cite{chruscinski}
\begin{eqnarray}\label{free}
\Phi_0 = \Phi (\pi/\omega_0)=\pi[1- \cos(2 \theta)],  \quad   \mbox{mod}  (2\pi),
\end{eqnarray}
which is a monotonic function of the mixing angle $\theta$.   This case  can serve  as a  reference  only for studying the  influence of the matter and dissipation.
For the dissipationless case but when the neutrino propagates through matter ($V_0\ne 0$) the dynamics is still unitary and  the analytic formula for GP  reads
\begin{eqnarray}\label{fazt}
\Phi(t) =  \mbox{Arg} \left[M(t) +i R(t)\right],
\end{eqnarray}
where
\begin{eqnarray}
M(t)&=&\cos(\Omega t) \cos(\omega t) + \frac{\omega}{\Omega} \sin(\Omega t) \sin(\omega t), \nonumber\\
R(t)&=&\cos(\Omega t) \sin(\omega t) - \frac{\omega}{\Omega} \sin(\Omega t) \cos(\omega t),
\end{eqnarray}
\begin{eqnarray} \label{omeg}
\Omega &=& \omega_0 \sqrt{1- 2 V \cos(2 \theta) + V^2}, \nonumber\\
 \omega & =& \omega_0 [ V - \cos(2\theta)], \quad
V =  V_0/2 \hbar \omega_0.
\end{eqnarray}
For this cyclic evolution with the period $T=\pi/\Omega$ one obtains
\begin{eqnarray}\label{faz}
\Phi(\pi/\Omega)=\pi \left[1- \frac{\cos(2 \theta) - V}{\sqrt{1- 2 V \cos(2 \theta)
+ V^2}}\right].
\end{eqnarray}
For the dimensionless potential parameter $V <<1$ we may Taylor expand the right-hand side of Eq. (\ref{faz}) and  obtain to the first order in $V$:
\begin{eqnarray}\label{smal}
\!\!\!\!\!\!\!\! \Phi(\pi/\Omega) = \Phi_{0} \! \left[ 1 +  V (1 +  \cos(2 \theta))  \right] \!\!, \;
\end{eqnarray}
where $\Phi_0$ is the reference GP in
Eq.(\ref{free}).
It follows that in the case of  neutrino propagation through
matter its GP increases in comparison to the vacuum reference curve. 
\begin{figure}[b]
\begin{center}
\includegraphics[width=0.25\textwidth,height=7cm,angle=270]{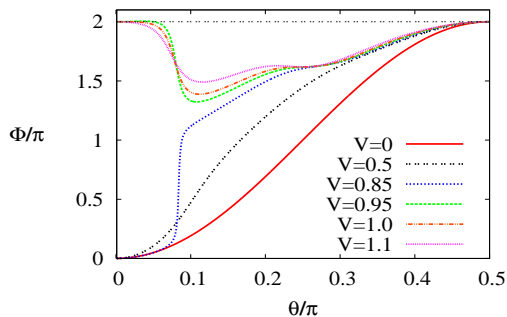}
\end{center}
\vspace{-4mm}
\caption{\label{fig1} (color online). The geometric phase $\Phi=\Phi(\pi/\omega_0)$ vs the mixing angle
$\theta$  for the neutrino interacting with an ordinary matter for selected values of the  dimensionless potential constant  $V$  (the case of absence of dissipation,  $C_{ij}=0$).
}
\end{figure}
The main note is that in  the absence of dissipation the GP does {\it not} depend on the Majorana angle $\phi$ and
therefore the GP
cannot be a tool for solving the   Dirac {\it vs}  Majorana neutrino dilemma.  In Fig. 1 we illustrate details of the influence of the neutrino-matter interaction on the GP at time $t=\pi/\omega_0$, i.e. $\Phi(\pi/\omega_0)$.  With the increase of the neutrino energy $E$, the
potential  $V$ rises and at  $V \approx 0.87$ the GP becomes a non--monotonic function of the mixing angle $\theta$, being significantly modified for $\theta$ smaller values. Yet with further increase of $V$ the GP as the function of $\theta$ stabilizes in the variation of $V$ and still at $V \approx 1.2$, i.e. when $\Delta^{2} m_{21}$ is active, it hardly feels the effect of further change of the neutrino energy, see Fig. 1.
Let us notice that at one period trip $t=\pi/\omega_{0}$ and both $\theta$ and $\Delta^{2}m_{21}$ equal to the experimental solar neutrino values $\theta_{\odot} = 0.188\, \pi$
and $\Delta^{2}m_{21} = 8.0 \times 10^{-5}$ ${\rm eV}^{2}$, respectively  \cite{Giunti-Kim},  the phase difference $\Phi(\pi/\omega_{0}) - \Phi_{0}$ becomes approximately equal to the geometric value  $\pi$ for $V \sim 1$ that means for the solar neutrino energies.
 Yet in the presence of ordinary matter the neutrino  evolution is no more  strictly cyclic at $t=\pi/\omega_0$  but
at $t=\pi/\Omega$. One can attempt to quantify to what extent the cyclic character of the evolution is affected by the interaction with an ordinary matter in terms of the trace distance between the state at $t=\pi/\omega_0$ and the initial state \cite{nielsen}:
\begin{figure}[b]
\vspace{-2mm}
\begin{center}
\includegraphics[width=0.25\textwidth,height=7cm,angle=270]{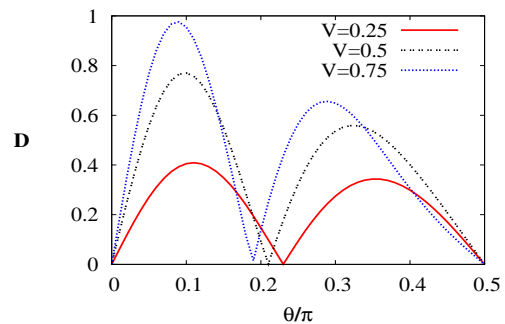}
\end{center}
\vspace{-4mm}
\caption{\label{fig2} (color online). The trace distance $D$ calculated as in Eq.(\ref{trejs}) plotted vs the mixing angle $\theta$ for different values of $V$. Remaining parameters are as in Fig. \ref{fig1}.}
\end{figure}
\begin{eqnarray}\label{trejs}
D=\frac{1}{2}||\rho(t=0)-\rho(t=\pi/\omega_0)|| \, , 
\end{eqnarray}
where the norm $||\varrho||=\mbox{Tr}\sqrt{\varrho^\dagger\varrho}$.
For a cyclic evolution, when the final and initial wave functions differ up to an overall phase  factor, $D=0$.
 As seen from Fig.~\ref{fig2}, the departure from the cyclic evolution quantified by $D$ is strongly affected by the mixing angle and, for certain angles the evolution, despite the presence of an ordinary matter ($V\neq 0$) remains  cyclic,
and this happens for $V\sim~1$ at the solar experimental value
$\theta = 0.188 \, \pi$ again.
It is interesting to analyze how the GP depends on choice of the time $t$. We have compared $\Phi(\pi/\omega_0)$ and $\Phi(\pi/\Omega)$
for the  mixing angle corresponding to the solar neutrino  value
$\theta =0.188\,\pi$.
For $V\in [0, 1.2]$, the difference $\Phi(\pi/\Omega) - \Phi(\pi/\omega_0)$  is extremely small and from the experimental point of view   negligible.  

\begin{figure}[t]
\begin{center}
\includegraphics[width=0.25\textwidth,height=7cm,angle=270]{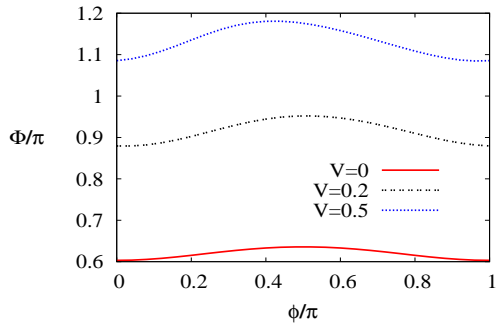}
\end{center}
\vspace{-4mm}
\caption{\label{fig3} (color online). The geometric phase $\Phi=\Phi(\pi/\omega_0)$ vs the Majorana angle $\phi$  in the presence of dissipation, $C(1,1)=C(2,2)=C(3,3)=1$,  the off--diagonal  contribution  $C(1,2)=C(2,1)=1/10$ and   for several values of $V$.
The solar neutrino mixing angle is
$\theta=0.188 \pi = 33.9^{\circ}$ \cite{Giunti-Kim}. }
\end{figure}

When the quantum system interacts with an environment, properties of the  GP can be radically modified \cite{my}. 
In the presence of dissipation, when the dissipation matrix  $C_{ij}$ in Eq.(\ref{dys}) is not identically zero, the GP can be determined by solving Eq.~(\ref{mast}) with (\ref{dys}) using, e.g. the
Bloch vector  formalism \cite{Bloch} to obtain the coupled evolution equations for
mean values $\langle \sigma_k(t)\rangle, k=x, y, z$. Next,  the
reduced  density matrix is found as  $\rho(t) = (1/2)[1 +\langle
\sigma_x(t)\rangle \sigma_x + \langle \sigma_y(t)\rangle \sigma_y +
\langle \sigma_z(t)\rangle \sigma_z]$. From this form one can obtain
the spectral decomposition (\ref{spect}) and the phase  $\Phi^{\alpha}(t)$.
Such an analytical form of the GP is, however,
rather cumbersome without exhibiting  much physical insight. Therefore, we  present here
the numerical results for the GP. The analysis has shown that none of the features of the GP is affected by the dissipative effects given by the diagonal matrix $C_{ij}\sim\delta_{ij}$.
 Hence all the results presented so far hold true for quite a general class of dissipative effects. It does {\it not} mean that the dynamics of neutrinos is unaffected by environmental noise as in the diagonal case  the trace distance $D$ in Eq.(\ref{trejs}) approaches constant value $D\approx 1/2$ with no regard to any choice of an initial preparation.
It is no more the case when there is an off--diagonal contribution to the dissipation  matrix $C_{ij}$.  In Fig. \ref{fig3} we present how the dissipation can affect the GP provided that there are nonvanishing  off--diagonal elements in $C_{ij}$.  The significant impact of dissipation is present only in a relatively narrow range of mixing angle $\theta$. Additionally, there is a feature which makes an off--diagonal dissipation worth studying: there is a nontrivial $\phi$--dependence of the GP.  In Fig.~\ref{fig3} the considerations are limited to a single nonvanishing element $C(1,2)=C(2,1)=1/10$ and for the solar neutrino mixing angle
$\theta=0.188 \pi = 33.9^{\circ}$. The results of other calculations, not reproduced here, show that the presented behavior is qualitatively generic.
A  most intriguing behavior on the role of the Majorana angle
emerges when  the angle $\phi$   is
allowed to vary and the mixing angle $\theta$ is fixed. One can observe that    the GP does depend on the Majorana angle $\phi$ in a non-monotonic way and  is always minimal for the Dirac neutrino; for the Majorana neutrino the GP is greater than for the Dirac neutrino. This property of the GP can provide a significant  test for the type of neutrino, the Majorana or Dirac one. Let us notice that this  effect originates essentially from  the dissipative character of an evolution since it is also  present for  the case $V=0$.  Since dissipation is a generic feature of the quantum world,  the $\phi$--dependence of the geometric phase seems to be  generic as well.  
%

In summary, the results reported in this paper show that
the GP can be a potentially useful indicator of various
properties of neutrinos and their environment. In the dissipationless
case, the GP does not depend on the Majorana
angle. However, in the presence of dissipation it is not the
case anymore: the GP does depend on the Majorana angle
and therefore can serve as a tool for determining the nature
of the Dirac vs the Majorana neutrino. The theoretical
analysis presented in the paper suggesting potential usefulness
of the GP as a tool for distinguishing neutrino type
achieves a real status of being useful provided that one can
perform an experiment measuring the GP in neutrino oscillations.
Any proposal of such an experiment, which
requires highly sophisticated experimental methods even
in the case on NMR-type systems \cite{fazmeasure}, is beyond the scope
of this brief report.\\

This work was supported by the MNiSW under Grant
No. N202 064936.

\end{document}